# S&P 500 Trend Prediction

**Group Members:** Shasha Yu, Qinchen Zhang, Yuwei Zhao

## Abstract

This project aims to predict short-term and long-term upward trends in the S&P 500 index using machine learning models and feature engineering based on the "101 Formulaic Alphas" methodology. The study employed multiple models, including Logistic Regression, Decision Trees, Random Forests, Neural Networks, K-Nearest Neighbors (KNN), and XGBoost, to identify market trends from historical stock data collected from Yahoo! Finance. Data preprocessing involved handling missing values, standardization, and iterative feature selection to ensure relevance and variability.

For short-term predictions, KNN emerged as the most effective model, delivering robust performance with high recall for upward trends, while for long-term forecasts, XGBoost demonstrated the highest accuracy and AUC scores after hyperparameter tuning and class imbalance adjustments using SMOTE. Feature importance analysis highlighted the dominance of momentum-based and volume-related indicators in driving predictions. However, models exhibited limitations such as overfitting and low recall for positive market movements, particularly in imbalanced datasets.

The study concludes that KNN is ideal for short-term alerts, whereas XGBoost is better suited for long-term trend forecasting. Future enhancements could include advanced architectures like Long Short-Term Memory (LSTM) networks and further feature refinement to improve precision and generalizability. These findings contribute to developing reliable machine learning tools for market trend prediction and investment decision-making.

# 1. Introduction

Stock markets play a vital role in facilitating efficient price discovery and enabling transactions by allowing buyers and sellers to exchange equity shares. Investors aim to achieve capital gains by accurately predicting stock price movements, which can involve buying at lower prices and selling at higher prices. This project sought to predict short-term upward trends in the S&P 500 index using machine learning models. Our approach was built on the "101 Formulaic Alphas" methodology introduced by Kakushadze in 2016, which provided a mathematical framework for constructing features that capture underlying market mechanisms. These alphas were used to develop predictive models, including Logistic Regression, Decision Trees, Random Forests, Neural Networks, and K-Nearest Neighbors (KNN). The primary objective was to evaluate these models and identify the most effective approach for forecasting market trends and enhancing investment strategies.

# 2. Data Collection and Preprocessing

The dataset used in this project was sourced from Yahoo! Finance and covered a timeframe from November 1, 2013, to October 31, 2024. The data included daily updates for the S&P 500 index and its 500 constituent stocks, capturing attributes such as Date, Open, High, Low, Close, Volume, and Adjusted Close. Given the necessity of rolling window calculations and the presence of missing entries in earlier data, the first sixteen months of data were excluded from the analysis. This adjustment ensured a robust and complete dataset for model training and prediction.

To prepare the data for analysis, missing values were forward-filled to maintain continuity and prevent gaps in the time series. Additionally, the final day's data was excluded to align the dataset with the lag introduced by predictive modeling. Standardization techniques were applied to ensure uniform scaling of features, thereby avoiding potential biases caused by differences in variable magnitudes.

Features were constructed based on the "101 Formulaic Alphas" methodology. These features encompassed various dimensions, such as momentum-based indicators, mean-reversion signals, volume-related metrics, and statistical factors. We first conducted feature dichotomy—if the number of unique values is less than 10, we considered it a discrete/categorical feature, otherwise continuous feature. Continuous features with a high rate of duplication—specifically those with over 20% identical values—were excluded to preserve temporal variability. Furthermore, features with high correlation (≥0.99) were iteratively removed to eliminate redundancy, resulting in a refined dataset comprising forty alphas for subsequent analysis.

# 3. Modeling Approach

The predictive models implemented in this project included Logistic Regression, Decision Trees, Random Forests, Neural Networks, and K-Nearest Neighbors. These models were chosen for their complementary strengths in handling classification tasks and their ability to address both parametric and non-parametric relationships within the data.

The response variable was designed to capture two types of trends. For short-term predictions, an upward trend was identified if the percentage change in the typical price exceeded 0.1%, which provided a practical early alert mechanism. For long-term predictions, an upward trend was defined based on whether the percentage change surpassed the 75th percentile of the previous sixty days' returns, offering insights into sustained market movements.

To ensure reproducibility and robust evaluation, a random seed of 42 was used for all model training processes. Hyperparameter tuning was conducted using five-fold cross-validation, which allowed the models to optimize their configurations for the best predictive performance.

# 4. Results and Analysis

## Short term prediction

**Logistic Regression**

LR output revealed balanced performance across classes, with an overall accuracy of 62% for training and 61% for testing. Precision, recall, and F1-scores were slightly higher for class 0.0 (no upward trend) compared to class 1.0 (upward trend). This indicated the model's tendency to favor the majority class (0.0), as reflected in its slightly better metrics for identifying the absence of upward trends. The close alignment between training and testing performance metrics suggested that the model generalized well to unseen data. However, the relatively low precision and recall for the upward trend class indicated room for improvement in correctly predicting positive market movements.

```
Classification Report for Logistic Regression: Training
              precision    recall  f1-score   support

         0.0       0.64      0.64      0.64      1042
         1.0       0.61      0.60      0.61       970

    accuracy                           0.62      2012
   macro avg       0.62      0.62      0.62      2012
weighted avg       0.62      0.62      0.62      2012

Classification Report for Logistic Regression: Testing
              precision    recall  f1-score   support

         0.0       0.63      0.60      0.62       265
         1.0       0.58      0.62      0.60       239

    accuracy                           0.61       504
   macro avg       0.61      0.61      0.61       504
weighted avg       0.61      0.61      0.61       504
```

## Decision Tree

The decision tree model was developed by optimizing hyperparameters through a randomized search process. Parameters such as maximum depth, minimum samples for splitting, and minimum samples per leaf were tuned using cross-validation to improve the model's generalization ability. The model was trained on a subset of the dataset to predict upward and downward trends in the S&P 500 index.

```
Best Hyperparameters: {'min_samples_split': 4, 'min_samples_leaf': 1, 'max_depth': 5, 'criterion': 'gini'}
Best Score: 0.540089420423128
```

Once trained, the decision tree was evaluated on both the training and testing datasets. The evaluation revealed that the model achieved a training accuracy of 68% and a testing accuracy of 59%, indicating moderate performance and some overfitting. Precision, recall, and F1-scores varied between the two classes, with slightly better results for identifying the absence of upward trends (class 0).

The model's predictions were further analyzed using a confusion matrix, which showed a relatively balanced distribution of true positives and false positives. However, the model faced challenges in fully capturing the complexities of upward trend prediction, as reflected in its lower recall for the upward trend class. This highlighted the potential need for ensemble techniques or feature refinement to enhance predictive accuracy.

```
 Classification Report for Decision Tree: Training
               precision    recall  f1-score   support

         0.0       0.70      0.67      0.68      1042
         1.0       0.66      0.69      0.67       970

    accuracy                           0.68      2012
   macro avg       0.68      0.68      0.68      2012
weighted avg       0.68      0.68      0.68      2012

 Classification Report for Decision Tree: Testing
               precision    recall  f1-score   support

         0.0       0.63      0.55      0.59       265
         1.0       0.56      0.63      0.59       239

    accuracy                           0.59       504
   macro avg       0.59      0.59      0.59       504
weighted avg       0.60      0.59      0.59       504
```

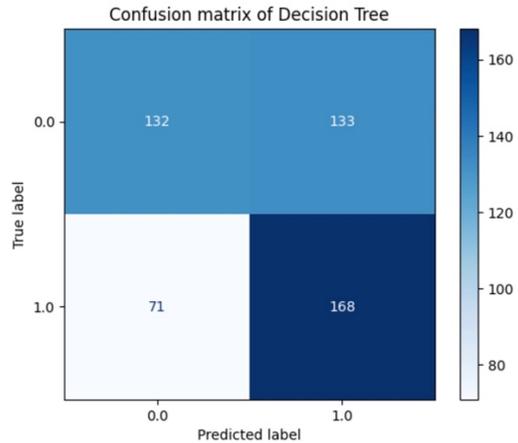

## Random Forest

The Random Forest model underwent a hyperparameter tuning process using randomized search to optimize its performance. The search involved testing various combinations of tree depth, minimum samples for splitting and leaf nodes, and the number of estimators to identify the configuration that achieved the highest F1-score during cross-validation.

```
Best Hyperparameters: {'n_estimators': 100, 'min_samples_split': 4, 'min_samples_leaf': 4, 'max_depth': 5}
Best Score: 0.583108288356945
```

After determining the optimal parameters, the model was trained and evaluated on both the training and testing datasets. The model achieved strong training performance, with an accuracy of 71% and balanced precision and recall scores, indicating its ability to capture patterns in the data. However, on the testing dataset, the accuracy dropped to 62%, highlighting some generalization issues. The model demonstrated relatively better performance in detecting upward trends (class 1) compared to other classifiers, but misclassifications persisted, as evidenced by the confusion matrix.

```
Classification Report for Random Forest: Training
               precision    recall  f1-score   support

         0.0       0.70      0.76      0.73      1042
         1.0       0.72      0.65      0.68       970

    accuracy                           0.71      2012
   macro avg       0.71      0.70      0.70      2012
weighted avg       0.71      0.71      0.71      2012

Classification Report for Random Forest: Testing
               precision    recall  f1-score   support

         0.0       0.64      0.66      0.65       265
         1.0       0.61      0.59      0.60       239

    accuracy                           0.62       504
   macro avg       0.62      0.62      0.62       504
weighted avg       0.62      0.62      0.62       504
```

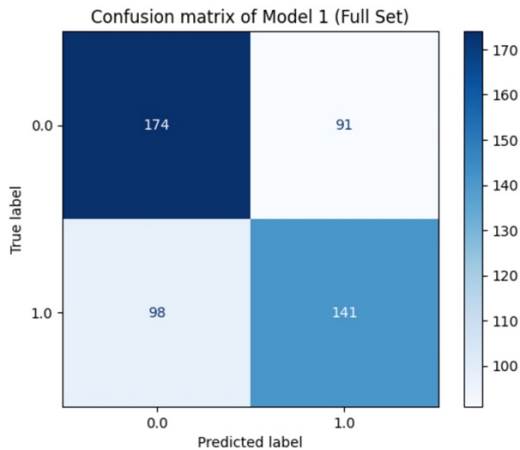

The feature importance analysis revealed the most influential predictors, with specific alphas, such as alpha054 and alpha053, contributing significantly to the model's decisions. A subset of 24 features with relative importance higher than 0.02 was extracted, suggesting that further refinement of input features could improve performance and reduce computational complexity. Overall, the Random Forest model showed promise with its ensemble-based approach but highlighted the need for additional adjustments to address overfitting and enhance generalization to unseen data.

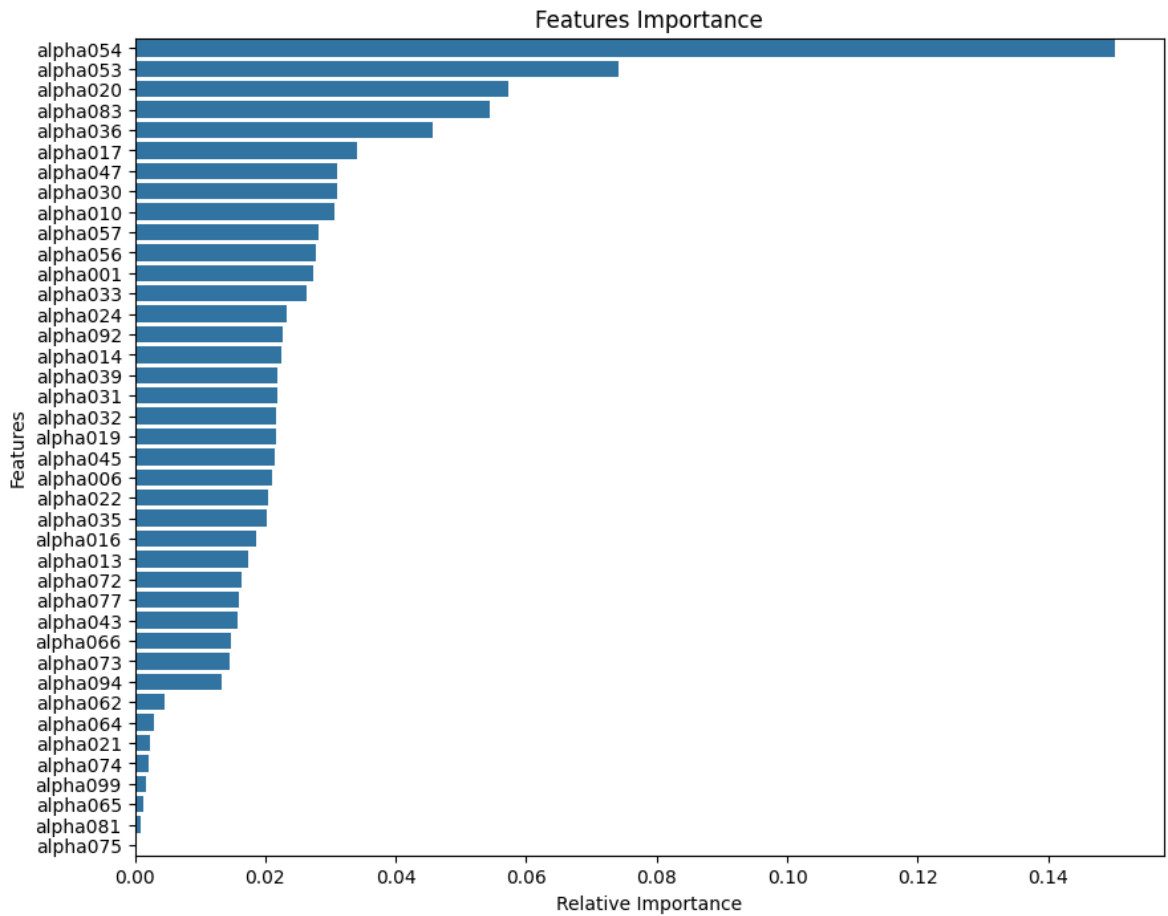

## KNN

The K-Nearest Neighbors (KNN) model was optimized by systematically evaluating various values for the number of neighbors (k) within a defined range. Each candidate value for k was tested, and the corresponding out-of-sample accuracy was recorded to identify the value that maximized performance. The selected k-value was then used to train the final KNN model on the training dataset.

The trained KNN model was evaluated using classification metrics, including precision, recall, and F1-score, for both training and testing datasets. The testing results indicated an overall accuracy of 61%, with balanced but moderate precision and recall values across the two classes. Specifically, the model demonstrated slightly higher recall for identifying the absence of upward trends (class 0), suggesting its effectiveness in detecting such instances. However, the recall for identifying upward trends (class 1) was lower, highlighting challenges in correctly capturing all positive trend instances.

The confusion matrix analysis revealed the distribution of correct and incorrect predictions, showing a notable number of misclassifications for upward trends. This emphasized the trade-offs inherent in the model's predictive performance and suggested potential areas for improvement, such as feature enhancement or integrating ensemble techniques.

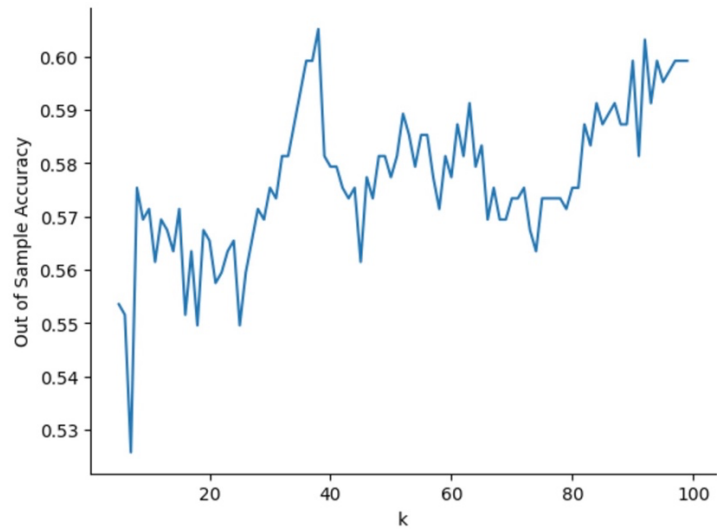

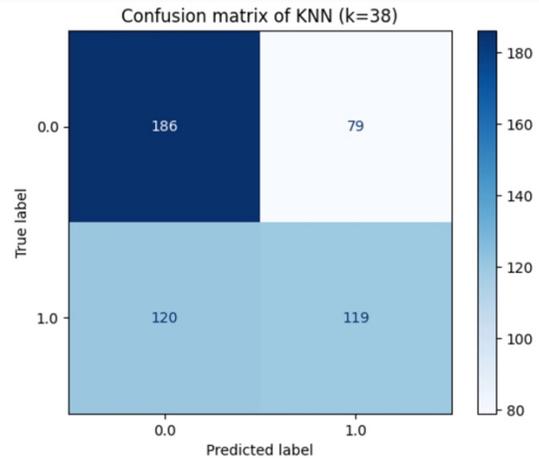

## XGB

The XGBoost model was optimized through a hyperparameter tuning process using a randomized search. Key parameters, including the learning rate, maximum tree depth, and number of estimators, were systematically tested to identify the configuration that maximized the model's F1-score. Subsampling ratios, column sampling, and gamma values were also considered to improve regularization and reduce overfitting.

Once the optimal hyperparameters were determined, the model was trained on the dataset, with class imbalance addressed by scaling the positive class weight. The evaluation revealed that the model achieved high accuracy on the training dataset, indicating that it effectively captured the underlying patterns. However, the testing results showed a moderate overall accuracy of 59%, suggesting some performance degradation when applied to unseen data.

The model performed slightly better in identifying the absence of upward trends (class 0), as reflected in higher precision and recall for this class compared to upward trends (class 1). Despite its robust training performance, the gap in testing results highlighted potential overfitting, suggesting that further adjustments, such as additional regularization or feature refinement, could enhance generalization. Overall, the XGBoost model demonstrated strong potential, particularly in scenarios where high precision for class 0 is critical.

```
Classification Report for XGBoost: Training
              precision    recall  f1-score   support

         0.0       0.78      0.77      0.77       948
         1.0       0.77      0.78      0.77       948

    accuracy                           0.77      1896
   macro avg       0.77      0.77      0.77      1896
weighted avg       0.77      0.77      0.77      1896

Classification Report for XGBoost: Testing
              precision    recall  f1-score   support

         0.0       0.64      0.55      0.59       246
         1.0       0.54      0.63      0.58       208

    accuracy                           0.59       454
   macro avg       0.59      0.59      0.59       454
weighted avg       0.59      0.59      0.59       454
```

For short-term predictions, the K-Nearest Neighbors (KNN) model emerged as the most effective approach. This model demonstrated strong accuracy and reliability in providing early alerts for upward trends. Key alphas contributing to its success included momentum-based indicators that highlighted relative price positioning during specific time periods, effectively capturing early signals of market movement.

## Long term prediction

The performance summary table and the corresponding ROC curve chart provided a comprehensive comparison of various machine learning models implemented for predicting long-term upward trends in the S&P 500 index. Among the models, XGBoost achieved the highest accuracy (72.02%) and the highest AUC score (0.6417), demonstrating its superior ability to balance true positive and false positive rates. Logistic Regression, with an AUC score of 0.6225, showed moderate performance but had relatively high recall (59.83%), indicating its strength in identifying true positives. On the other hand, KNN excelled in recall (65.57%) but had a low precision score, reflecting its tendency to generate false positives. Decision Tree and Random Forest models achieved decent accuracy but struggled with recall and AUC, while the Neural Network model showed balanced but suboptimal metrics across all categories. The ROC curve highlighted the trade-offs in predictive performance, with XGBoost and Random Forest models achieving better discrimination between classes compared to other approaches. These insights underlined the varying strengths of each model and the need to align model selection with specific business or analytical priorities.

|                        | accuracy | precision | recall   | f1-score | auc score |
|------------------------|----------|-----------|----------|----------|-----------|
| Logistic Regression    | 0.634921 | 0.350962  | 0.598361 | 0.442424 | 0.622479  |
| Decision Tree          | 0.625000 | 0.286624  | 0.368852 | 0.322581 | 0.537829  |
| Random Forest          | 0.700397 | 0.306667  | 0.188525 | 0.233503 | 0.526199  |
| Random Forest - Subset | 0.704365 | 0.324675  | 0.204918 | 0.251256 | 0.534396  |
| Neural network         | 0.662698 | 0.285714  | 0.262295 | 0.273504 | 0.526435  |
| KNN                    | 0.537698 | 0.295203  | 0.655738 | 0.407125 | 0.577869  |
| XGBoost                | 0.720238 | 0.382716  | 0.254098 | 0.305419 | 0.561604  |

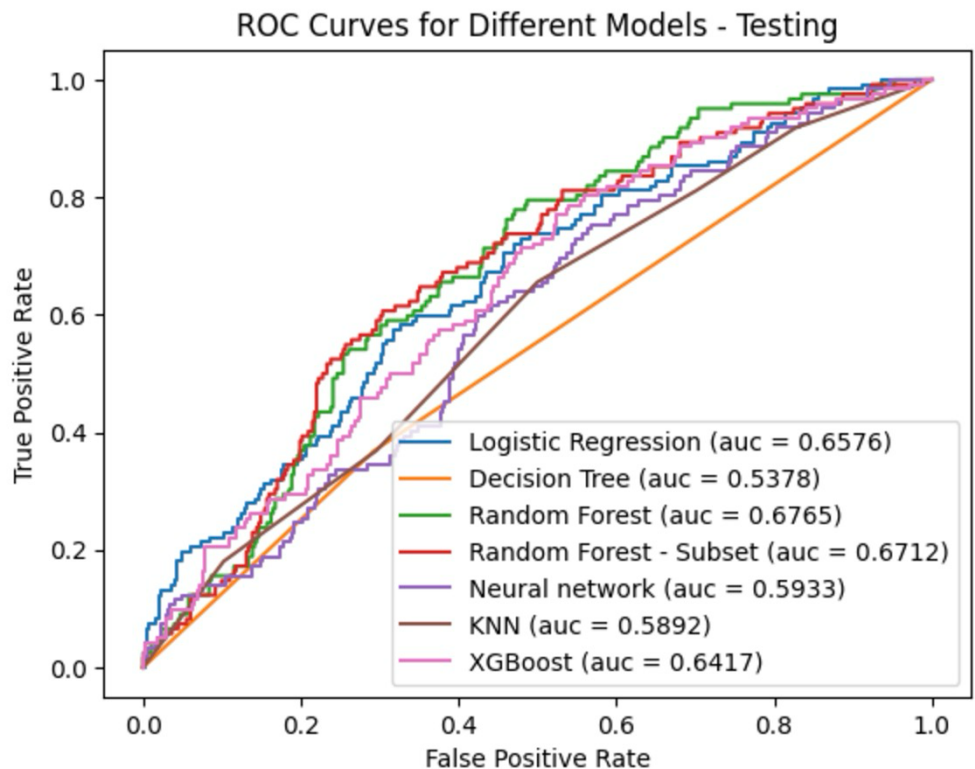

In the context of long-term predictions, the XGBoost model showed the greatest potential, particularly when applied to datasets balanced using the Synthetic Minority Oversampling Technique (SMOTE). The original dataset exhibited a significant imbalance, with 491 instances of upward trends compared to 1521 downward trends. This imbalance led models to disproportionately favor the majority class, undermining their ability to detect true positives. By applying SMOTE, the dataset was rebalanced, improving the precision and recall of the long-term models. The key alphas influencing long-term predictions emphasized weighted price positions and trading volumes, focusing on stocks with significant price movements and active trading.

Despite these results, it was observed that no single model consistently outperformed others across all metrics. Logistic Regression and Neural Networks offered moderate performance but struggled with recall, which limited their effectiveness in capturing true positive trends. Decision Trees and Random Forests demonstrated high accuracy but required careful hyperparameter tuning to avoid overfitting.

## Model Comparison

The comparison of models revealed distinct strengths and weaknesses. For short-term predictions, KNN provided robust and timely alerts, making it suitable for investors seeking quick decision-making tools. In contrast, XGBoost excelled in long-term trend forecasting but required additional optimization to address data imbalances and improve precision. Logistic Regression and Neural Networks, while reliable in certain aspects, were less effective in capturing nuanced market signals. Decision Trees and Random Forests offered high accuracy but exhibited sensitivity to hyperparameter configurations, which necessitated careful calibration.

## 5. Conclusion and Recommendations

The study concluded that KNN is the preferred model for short-term predictions due to its ability to deliver accurate early alerts, while XGBoost is recommended for long-term forecasting given its capacity to handle complex relationships within the data. However, limitations were identified, including overall low recall and precision, particularly in long-term predictions. Some models also showed signs of overfitting, indicating the need for further refinement.

To address these challenges, future research could explore more advanced architectures such as Long Short-Term Memory (LSTM) networks, which are well-suited for time-series data. Additionally, techniques to better handle class imbalances, such as advanced resampling methods or cost-sensitive learning, could enhance model performance. Applying the framework at the company level rather than the index level may also yield more granular and interpretable results, offering deeper insights for targeted investment strategies.

## 6. Acknowledgments

We express our gratitude to Kakushadze for the "101 Formulaic Alphas" methodology, which served as the foundation for this project. We also acknowledge the contributions of all team members in data analysis, model development, and reporting, which were instrumental in the successful completion of this study.